\documentclass[aps,prb,amsmath,amssymb,reprint,superscriptaddress,preprintnumbers,showpacs,intlimits]{revtex4-1}
\pdfoutput=1
\usepackage{bm,latexsym,mathrsfs,enumerate,color}
\usepackage[mathcal]{euscript}
\usepackage[breaklinks=true,unicode=true,urlcolor = blue,colorlinks = true,citecolor = blue,linkcolor = blue]{hyperref}
\usepackage{graphicx}
\usepackage[normalem]{ulem}

\renewcommand{\vec}[1]{\bm{#1}}
\begin{document}

\title{Periodic magnetization structures generated by transverse spin-current in magnetic nanowires}

\author{Volodymyr P. Kravchuk}
 \email{vkravchuk@bitp.kiev.ua}
 \affiliation{Bogolyubov Institute for Theoretical Physics, 03143 Kiev, Ukraine}

\author{Oleksii M. Volkov}
\affiliation{Taras Shevchenko National University of Kiev, 01601 Kiev, Ukraine}

\author{Denis D. Sheka}
\affiliation{Taras Shevchenko National University of Kiev, 01601 Kiev, Ukraine}

\author{Yuri Gaididei}
 \affiliation{Bogolyubov Institute for Theoretical Physics, 03143 Kiev, Ukraine}

\date{\today}

%
%

\begin{abstract}
Magnetization behavior of long nanowires with square cross-section under influence of strong perpendicular spin-polarized current is studied theoretically. The study is based on Landau-Lifshitz-Slonczewski phenomenology. In the no current case the wire is magnetized uniformly along its axis. For small currents the wire magnetization remains uniform but it inclines with respect to the wire axis within the plane perpendicular to the current direction. With the current increasing the inclination angle increases up to the maximum value $\pi/4$. Further current increase leads either to saturation or to stable periodic multidomain structure depending on the wire thickness. For thick wires a hysteresis is observed in the saturation process under the action of current. All critical parameters of the current induced magnetization behavior are determined theoretically. The study is carried out both analytically and using micromagnetic simulations.
\end{abstract}




\maketitle

\section{Introduction}
\label{sec:intro}
The usage of spin-polarized current is a convenient way to control the magnetic structures in nanowires without applying the external magnetic fields, that enables increasing of density of arrays of nanoscale elements in purely current controlled devices\cite{Parkin08,Allwood05}. Usually, the current is passed along the magnetic wire, this is so-called CIP (current-in-plane) configuration. In this case the influence of the current on the dynamics of domain walls is widely studied both theoretically and experimentally, see reviews [\onlinecite{Lindner10,Tatara08,Marrows05,Klaui08}].

In the last few years an interest to CPP (current perpendicular to the plane) configuration of nanowires appears, e.g. it was shown theoretically\cite{Khvalkovskiy09} that in CPP case the velocity of the domain wall can be much higher than those for a CIP stripe with similar applied current densities. The similar domain wall motion was recently observed experimentally\cite{Boone10a,Chanthbouala11} characteristic currents densities are much smaller than the one commonly used in CIP configurations. It was also shown\cite{YanEPL10,BoonePRL10} that the domain wall can move even faster for the time alternating current in CPP wire structure.

In this work we also consider the CPP nanowire configuration, see Fig.~\ref{fig:coords}. We study possible stationary states and their stability under the current action. We found that there exists a critical transverse size $h_c$ of the nanowire. For sizes $h<h_c$ a stable periodic domain structure appears in the pre-saturated regime, for thicker wires $h>h_c$ the process of transverse saturation by current has a hysteresis, i.e. the critical current of transition to the saturated state (in the process of the current increasing) is larger than the critical current of the saturated state breaking (in the process of the current decreasing). Recently we reported on the formation of stable periodic square vortex-antivortex lattice (vortex crystal) in the pre-saturated regime of the magnetic films of CPP configuration\cite{Volkov11,Gaididei12a}. The idea of this paper is to demonstrate that strong restriction of film size in one dimension significantly affects the current-induced behavior of magnetization. In addition to the above-mentioned results we found that only stationary states take place in the nanowire with CPP, contrary to films where the essentially dynamic regimes, such as fluid-like and gas-like dynamics of vortices and antivortices appear for current decreasing\cite{Volkov11,Gaididei12a}.


The paper is organized in the following way: in Section~\ref{sec:model} we introduce a mathematical model of the one-dimensional magnetic nanowire which is based on the Landau-Lifshitz-Slonczewski equation. Here we adapt the spin operators representations of Holstein-Primakoff and Tyablikov for our classical system. Considering a case of soft ferromagnet we take into account exchange and dipole-dipole interactions and derive the corresponding Hamiltonian in the wave-vector space. In Section~\ref{sec:Stat-states} we obtain two possible uniform stationary solutions and analyse their stability. The nonuniform stationary solution -- periodic domain structure, which arises as a result of instability of the saturates state, we study in the Section~\ref{sec:domains}. In whole, the magnetization behavior of the nanowire under the current influence is summarized in the Conclusions and mathematical details of deriving of an equation of motion and Hamiltonian are placed in two Appendixes.


\section{One-dimensional discrete model}\label{sec:model}
\begin{figure}
\includegraphics[width=0.4\columnwidth]{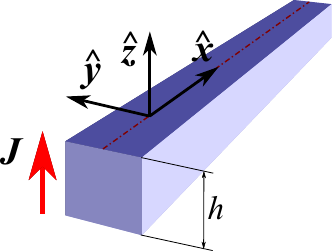}
\caption{Geometry and notations of the problem. A long nanowire of square cross-section and thickness $h$. The spin-polarized current $J$ flows perpendicularly to the wire.}\label{fig:coords}
\end{figure}

We consider here a narrow nanowire of square cross-section whose transverse size $h$ is small enough to ensure one-dimensionality of the magnetization, see Fig.~\ref{fig:coords}. In other words we assume that the magnetization is varying only along the wire -- along $\hat{\vec{x}}$-axis. This assumption works for the cases when $h$ is comparable or smaller than the characteristic magnetic length of the material, see bellow Section~\ref{sec:hamiltonian}. Total length of the wire $L\gg h$. The frame of reference is chosen as shown in the Fig.~\ref{fig:coords}.
The magnetic media is modeled as a discrete cubic lattice of magnetic moments $\vec M_{\vec\nu}$, where $\vec\nu=a(\nu_x,\,\nu_y,\,\nu_z)$ is three dimensional index with $a$ being the lattice constant and $\nu_x,\,\nu_y,\,\nu_z\in\mathbb{Z}$. It is convenient to introduce the following notations: $\mathcal{N}_x=L/a$ is the total number of lattice nodes along $\hat{\vec x}$-axis and $\mathcal{N}_{s}=h^2/a^2$ is the number of nodes within the cross-section square.

We base our study on the one-dimensional discrete Landau-Lifshitz-Slonczewski equation:\cite{Slonczewski96,Berger96,Slonczewski02}
\begin{equation} \label{eq:LLS}
\dot{\vec{m}}_{n} = \vec m_{n}\times{\partial\mathcal{E}}/{\partial\vec{m}_{n}}-\mathrm{j}\varepsilon_n \vec m_{n}\times[\vec m_{n}\times\hat{\vec z}],
\end{equation}
which describes the magnetization dynamics under influence of spin-polarized current which flows perpendicularly to the wire along $\hat{\vec z}$-axis as it is shown in the Fig.~\ref{fig:coords}. It is also assumed that the current flow and its spin-polarization are of the same direction in \eqref{eq:LLS}. The index $n=a\nu_x$ numerates the normalized magnetic moments $\vec m_{n} = \vec M_{ n}/|\vec M_{n}|$ along the wire axis. The overdot indicates derivative with respect to the rescaled time in units of $(4\pi\gamma_0 M_s)^{-1}$, $\gamma_0$ is gyromagnetic ratio, $M_s$ is the saturation magnetization, and $\mathcal{E}=E/(4\pi M_s^2 a^3 \mathcal{N}_s)$ is dimensionless magnetic energy. The normalized electrical current density $\mathrm{j}=J/J_0$, where $J_0=4\pi M_s^2|e|h/\hbar$ with $e$ being electron charge and $\hbar$ being Planck constant.  The spin-transfer torque efficiency function $\varepsilon_n$ has the form
\begin{equation}\label{eq:effic-func}
\varepsilon_n=\frac{\eta}{(1+\Lambda^{-2})+(1-\Lambda^{-2})m_n^z},
\end{equation}
where $\eta$ is the degree of spin polarization and parameter $\Lambda\geqslant1$ describes the mismatch between spacer and ferromagnet resistance \cite{Slonczewski02,Sluka11}.

\subsection{Holstein-Primakoff-Tyablikov representation}

For the future analysis it is convenient to introduce complex amplitude $\psi_n$ of deviation from the stationary distribution of the magnetization (possible stationary states are considered bellow). For this purpose, we use classical analogue of Holstein-Primakoff representation\cite{Holstein40} for spin operators. Here we consider general case of an arbitrary form of the stationary state, where the magnetization distribution is determined by the unit vector
\begin{equation}
\vec\gamma_n=(\sin\Theta_n\cos\Phi_n,\,\sin\Theta_n\sin\Phi_n,\,\cos\Theta_n),
\end{equation}
which can be index (coordinate) dependent.
For this purpose, we also use the eigen-representation proposed by Tyablikov\cite{Tyablikov75} and finally, combining these two approaches, we obtain the following representation for the magnetization
\begin{subequations}\label{eq:tyablikov}
\begin{align}\label{eq:m-repres}
&\vec{m}_n=\vec{\gamma}_n\frac{1-|\psi_n|^2}{2} +\vec A_n\psi_n\sqrt{2-|\psi_n|^2}+c.c.,
\end{align}
where vector $\vec A_n$ is orthogonal to $\vec\gamma_n$ and has the following form
\begin{align}
\vec A_n =&\frac12\bigl(\cos\Theta_n\cos\Phi_n+i\sin\Phi_n,\\ \nonumber
&\cos\Theta_n\sin\Phi_n-i\cos\Phi_n,\,-\sin\Theta_n\bigr).
\end{align}
\end{subequations}

Using the representation \eqref{eq:tyablikov} we obtain the following expressions for the magnetization components
\begin{subequations}\label{eq:m-components}
\begin{align}
\label{eq:mx_imy}&m^x_n+im^y_n=e^{i\Phi_n}\biggl\{\sin\Theta_n\left(1-|\psi_n|^2\right)\\ \nonumber
&+\sqrt{2-|\psi_n|^2}\left[\psi_n\cos^2\frac{\Theta_n}{2}-\psi_n^*\sin^2\frac{\Theta_n}{2}\right]\biggr\},\\
\label{eq:mz}&m^z_n=\cos\Theta_n(1-|\psi_n|^2)-\frac{\sin\Theta_n}{2}\sqrt{2-|\psi_n|^2}(\psi_n+\psi_n^*).
\end{align}
\end{subequations}
It should be noted that for the case of absence of deviations we have $\psi_n=0$ and expressions \eqref{eq:m-components} result in the magnetization orientated along vector $\vec\gamma_n$.  The similar procedure we used in Ref.~\onlinecite{Gaididei12a} for the case of the uniform stationary state which is oriented along the current.

Substituting \eqref{eq:m-components} into \eqref{eq:LLS} enables one to proceed from the set of equations for the magnetization components to a single equation for complex valued function~$\psi$:
\begin{equation}\label{eq:eq-motion-psi}
-i\dot\psi_n=\frac{\partial\mathcal{E}}{\partial\psi_n^*}+\mathcal{F}^{st}_n,
\end{equation}
where the spin-torque term $\mathcal{F}^{st}_n$ has the following form
\begin{equation}\label{eq:Fst-psi-exact}
\begin{split}
\mathcal{F}_n^{st}=&i\mathrm{j}\varepsilon_n\Biggl[\cos\Theta_n\psi_n\left(1-\frac12|\psi_n|^2\right)\\
+&\sin\Theta_n\frac{1-\frac14(3-|\psi_n|^2)(\psi_n^2-|\psi_n|^2)}{\sqrt{2-|\psi_n|^2}}\Biggr].
\end{split}
\end{equation}
Here the efficiency function $\varepsilon_n$ depends on $\psi_n$ according to \eqref{eq:effic-func} and \eqref{eq:mz}. For details of deriving \eqref{eq:eq-motion-psi} see Appendix~\ref{ap:the-equation}.

Since the Eq.~\eqref{eq:eq-motion-psi} describes the deviation from a stationary solution, it has a very convenient form for analysis of stability of the given stationary state. Although the Eq.~\eqref{eq:eq-motion-psi} can be used for arbitrary stationary state, in what follows we restrict ourselves to the case of spatially uniform stationary states, so that $\Theta_n=\Theta=\mathrm{const}$ and $\Phi_n=\Phi=\mathrm{const}$. Here we study the linear stability, thus it is enough to use the linearized form of Eq.~\eqref{eq:eq-motion-psi}. It is also convenient to proceed to the wave-vector representation using the discrete Fourier transform
\begin{subequations}\label{eq:Fourier-def}
\begin{align}
\label{eq:four-inv}&\psi_{n}=\frac{1}{\sqrt{\mathcal{N}_{x}}}\sum\limits_{k}\hat\psi_{k}e^{i k n},\\
\label{eq:four}&\hat\psi_{k}=\frac{1}{\sqrt{\mathcal{N}_{x}}}\sum\limits_{n}\psi_{n}e^{-i k n}
\end{align}
with the orthogonality condition
\begin{align} \label{eq:orth-cond}
\sum\limits_{n}e^{i( k- k')n}=\mathcal{N}_{x}\Delta(k-k'),
\end{align}
\end{subequations}
where $k=\frac{2\pi}{L}l$ is two-dimensional discrete wave vector, $l\in\mathbb{Z}$, and $\Delta(k)$ is the Kronecker delta. Applying \eqref{eq:Fourier-def} to the linearized equation \eqref{eq:eq-motion-psi} one gets the following equation of motion in the wave-vector space
\begin{equation}\label{eq:motion-four}
\begin{split}
-i\dot{\hat{\psi}}_k=\frac{\partial\mathcal{E}^0}{\partial\hat{\psi}^*_k}&+i\mathrm{j}\varepsilon^0\biggl[\sqrt{\frac{\mathcal{N}_x}{2}}\sin\Theta+\cos\Theta\hat\psi_k\\
&+\frac{\varepsilon^0}{2\eta}\left(1-\Lambda^{-2}\right)\sin^2\Theta(\hat\psi_k+\hat\psi^*_{-k})\biggr],
\end{split}
\end{equation}
where $\varepsilon^0=\eta/\left[(1+\Lambda^{-2})+(1-\Lambda^{-2})\cos\Theta\right]$ is the spin-transfer torque efficiency function \eqref{eq:effic-func} for the case $\psi_n=0$ and $\mathcal{E}^0$ denotes the harmonic part of the normalized energy.

\subsection{Hamiltonian of the system}\label{sec:hamiltonian}
We consider here the case of soft ferromagnet, therefore we take into account only two contributions into the total energy: $E=E_{ex}+E_{d}$. Here
\begin{equation} \label{eq:Eex}
E_{ex}=-\mathcal{S}^2\mathcal{J}\sum\limits_{\vec\nu,\vec\delta}\vec m_{\vec\nu}\cdot\vec m_{\vec\nu+\vec\delta}
\end{equation}
is the exchange contribution, where $\vec\delta=a(\delta_x,\,\delta_y,\,\delta_z)$ is a three dimensional index which numerates the nearest neighbors of an atom, value of spin is denoted with $\mathcal{S}$ and $\mathcal{J}>0$ is exchange integral between two nearest atoms.

The other term is the dipole-dipole energy
\begin{equation} \label{eq:Ems}
E_\mathrm{d}=\frac{M_s^2a^6}{2}\!\!\sum\limits_{\vec\nu\ne\vec\mu}\biggl[\frac{ (\vec m_{\vec\nu}\!\cdot\! \vec m_{\vec\mu})}{r_{\vec\nu\vec\mu}^3}
-3\frac{\left(\vec m_{\vec\nu}\!\cdot\! \vec r_{\vec\nu\vec\mu}\right) \left(\vec m_{\vec\mu}\! \cdot\! \vec r_{\vec\nu\vec\mu}\right)}{r_{\vec\nu\vec\mu}^5}\biggr],
\end{equation}
where we introduce the notation $\vec r_{\vec\nu\vec\mu}=(x_{\vec\nu\vec\mu},\,y_{\vec\nu\vec\mu},\,z_{\vec\nu\vec\mu})=\vec\mu-\vec\nu$.

Hereinafter we will be interested only in the harmonic approximation $\mathcal{E}^0$ of the normalized energy which includes terms not higher than $\mathcal{O}(|\psi|^2)$. Thus
\begin{equation}\label{eq:E0}
\mathcal{E}^0=\mathcal{E}_{ex}^0+\mathcal{E}_d^0,
\end{equation}
where $\mathcal{E}_{ex}^0$ and $\mathcal{E}_d^0$ are harmonic parts of exchange and dipole-dipole energies respectively. Substituting now the representation \eqref{eq:m-components} to \eqref{eq:Eex} and applying the Fourier transform \eqref{eq:Fourier-def} one obtains that the harmonic approximation of the normalized exchange energy reads
\begin{equation}\label{eq:Eex-four}
\mathcal{E}_{ex}^0=\ell^2\sum\limits_k k^2|\hat\psi_k|^2,
\end{equation}
where the exchange length $\ell=\sqrt{\mathcal{S}^2\mathcal{J}/(2\pi M_s^2a)}$ determines the scale of magnetization inhomogeneities. Here we neglected possible surface effects which can arise due to the different number of the nearest neighbors at the surface. The derivation of \eqref{eq:Eex-four} is analogous to one presented in Appendix A1 of the Ref.~\onlinecite{Gaididei12a}.

Let us proceed now to the dipole-dipole contribution. Using the fact that the magnetization depends only on $x$-coordinate one can present the expression \eqref{eq:Ems} in form
\begin{equation}\label{eq:Ed-1D}
\begin{split}
&E_\mathrm{d}=\frac{M_s^2a^6}{2}\!\!\sum\limits_{\nu_x,\mu_x}\left[\sum\limits_{\varsigma=x,y,z}\mathcal{A}_{\nu_x\mu_x}^\varsigma m_{\nu_x}^\varsigma m_{\mu_x}^\varsigma+\mathcal{B}_{\nu_x\mu_x}m_{\nu_x}^y m_{\mu_x}^z\right],\\
&\mathcal{A}_{\nu_x\mu_x}^\varsigma=\sum\limits_{\begin{smallmatrix}\bar{\vec\nu},\bar{\vec\mu},\\ \vec\nu\ne\vec\mu\end{smallmatrix}}\frac{r_{\vec\nu\vec\mu}^2-3\varsigma_{\vec\nu\vec\mu}^2}{r_{\vec\nu\vec\mu}^5},\quad\mathcal{B}_{\nu_x\mu_x}=-6\sum\limits_{\begin{smallmatrix}\bar{\vec\nu},\bar{\vec\mu},\\ \vec\nu\ne\vec\mu\end{smallmatrix}}\frac{y_{\vec\nu\vec\mu}z_{\vec\nu\vec\mu}}{r_{\vec\nu\vec\mu}^5}.
\end{split}
\end{equation}
Here we introduced the notations $\bar{\vec\nu}=(\nu_x,\,\nu_y)$ and $\bar{\vec\mu}=(\mu_x,\,\mu_y)$ for the sake of simplicity.
It is easy to show\footnote{One should present $\mathcal{B}_{\nu_x\mu_x}$ in integral form and perform the straightforward integration.} that $B_{\nu_x\mu_x}\equiv0$ for the case of rectangular wire cross-section, the same is true for the cross-sections in form of disk (cylindrical wire) or ring (tube). Substituting \eqref{eq:m-components} into \eqref{eq:Ed-1D} and applying the Fourier transform \eqref{eq:Fourier-def} we obtain the following harmonic approximation for the normalized dipole-dipole energy

\begin{equation}\label{eq:Ed-four-square}
\begin{split}
&\mathcal{E}^0_d=-\sum\limits_k\biggl\{3\sqrt{2\mathcal{N}_x}\sin\Theta g(0)\cos\Phi(\cos\Theta\cos\Phi+i\sin\Phi)\hat\psi_k\\
&+\frac32 g(k)(\cos\Theta\cos\Phi+i\sin\Phi)^2\hat\psi_k\hat\psi_{-k}\\
&+\frac12\left[g(k)+2g(0)\right](1-3\sin^2\Theta\cos^2\Phi)|\hat\psi_k|^2+c.c.\biggr\},
\end{split}
\end{equation}
where the function $g(k)$ can be presented approximately as
\begin{equation}\label{eq:g-approx}
g(k)\approx\frac12\left[I_1\left(\frac{kh}{\sqrt\pi}\right)K_1\left(\frac{kh}{\sqrt\pi}\right)-\frac13\right],
\end{equation}
with $I_1(x)$ and $K_1(x)$ being modified Bessel functions of the first and second types perspectively. For the exact form of $g(k)$ and other details see Appendix~\ref{ap:dipole}.

\section{Stationary states of the system}\label{sec:Stat-states}
The stationary magnetization distribution is determined by Eq.~\eqref{eq:motion-four} with $\hat\psi\equiv0$
\begin{equation}\label{eq:ground}
\left.\frac{\partial\mathcal{E}^0}{\partial\hat{\psi}^*_k}\right|_{\hat\psi_k=0}+\sin\Theta\frac{i\mathrm{j}\varepsilon^0\sqrt{\mathcal{N}_x}}{\sqrt{2}}=0.
\end{equation}

\begin{figure}
\includegraphics[width=\columnwidth]{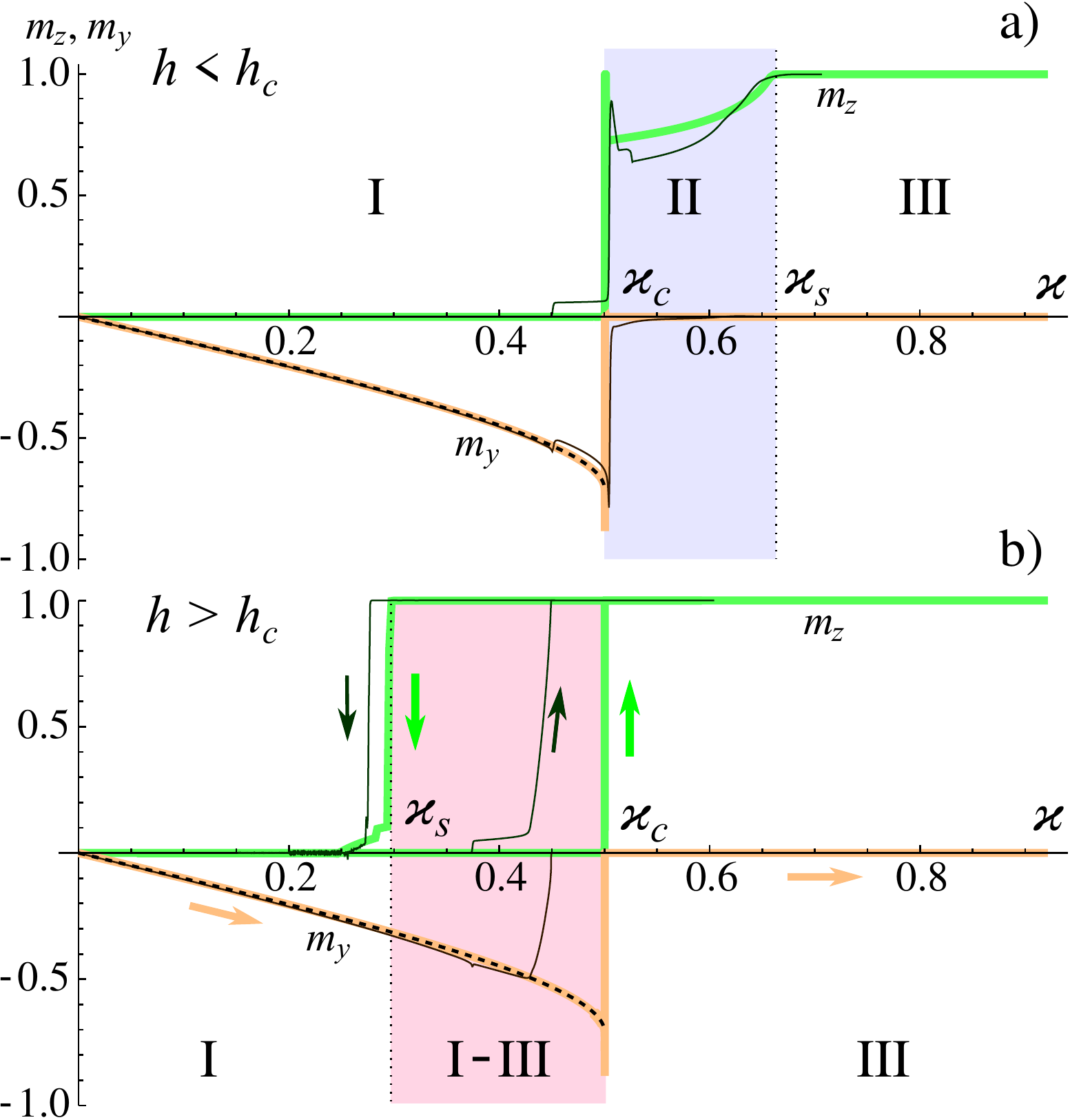}
\caption{Components of total magnetization of the nanowire as functions of the applied current. Panels a) and b) correspond to square permalloy nanowires with transverse sizes $h=5$~nm and $h=15$~nm respectively. The results of micromagnetic simulations are shown by solid lines: thick and thin lines corresponds to simulations with periodic boundary conditions along $\hat{\vec x}$-axis and to the wire of finite length $L=1\,\mu m$ respectively. The analytical solution \eqref{eq:my-ground} is shown by the dashed line. The current density $\varkappa$ is normalized in accordance with \eqref{eq:varkappa-def}. In region I the wire is magnetized uniformly within the plane perpendicular to the current flow and inclined at an angle $\Phi$ to the wire axis, see  \eqref{eq:low-current-sol} and Fig.~\ref{fig:diagram}b. The region II corresponds to a periodic domain structure, see Fig.~\ref{fig:domains}, and region III corresponds to a saturated state, see Fig.~\ref{fig:diagram}c. For the thick wires (panel b) the hysteresis region I-III arises, where the magnetization state coincides with the former state: either with the state I for the process of current increasing, or with the state III for the current decreasing from the saturation value. The hysteresis loop is shown only for component $m_z$ in order to not overload the figure.}\label{fig:mxy-vs-kappa}
\end{figure}

Substitution \eqref{eq:Ed-four-square} into \eqref{eq:ground} results in the equation for ground states of the system for a given value of the current~$\mathrm{j}$
\begin{equation}\label{eq:groung-angles}
\sin\Theta\left[i\mathrm{j}\varepsilon^0-\frac{\cos\Phi}{2}\left(\cos\Theta\cos\Phi-i\sin\Phi\right)\right]=0,
\end{equation}
where we took into account that $g(0)=1/12$.

Equation~\eqref{eq:groung-angles} has two solutions. The first one, namely
\begin{equation}\label{eq:satur-sol}
\sin\Theta=0,
\end{equation}
describes a saturated state when the wire is uniformly magnetized along $\hat{\vec z}$-axis, i.e. along the current direction. The second solution of Eq.~\eqref{eq:groung-angles} reads
\begin{subequations}\label{eq:low-current-sol}
\begin{align}
&\left\{\begin{aligned}
&\cos\Theta=0,\\
&\sin\Phi\cos\Phi=-\varkappa,
\end{aligned}\right.\\
\label{eq:varkappa-def}
&\varkappa=\mathrm{j}\frac{2\eta}{1+\Lambda^{-2}}.
\end{align}
\end{subequations}
In terms of magnetization components \eqref{eq:m-components} the solution \eqref{eq:low-current-sol} has the following form
\begin{subequations}\label{eq:m-low-current}
\begin{align}
&m^z_n=0,\\
\label{eq:mx-ground}&m^x_n(\varkappa)=\cos\Phi(\varkappa)=\left[\frac{1+\sqrt{1-4\varkappa^2}}{2}\right]^{\frac12},\\
\label{eq:my-ground}&m^y_n(\varkappa)=\sin\Phi(\varkappa)=-\left[\frac{1-\sqrt{1-4\varkappa^2}}{2}\right]^{\frac12}\approx-\varkappa.
\end{align}
\end{subequations}
The solution \eqref{eq:m-low-current} as well as \eqref{eq:low-current-sol} exists for the current interval $0\leq\varkappa\leq1/2$, that corresponds to varying of the angle $\Phi$ in interval $-\pi/4\leq\Phi\leq0$.

In the following we consider stability of each of the solutions \eqref{eq:satur-sol} and \eqref{eq:low-current-sol}. Let us start from the stability analysis of the solution \eqref{eq:low-current-sol}. To obtain the equation of motion linearized in vicinity of the stationary solution we substitute \eqref{eq:low-current-sol} into \eqref{eq:motion-four} and \eqref{eq:Ed-four-square}, that results in
\begin{subequations}\label{eq:Eq-motion-xy}
\begin{align}
\label{eq:Eq-motion-xy-Eq}&-i\dot{\hat\psi}_k=\frac{\partial\mathcal{E}^0}{\partial\hat\psi^*_k}+i\frac{\varkappa}{4}\frac{1-\Lambda^{-2}}{1+\Lambda^{-2}}\left(\hat\psi_k+\hat\psi_{-k}^*\right),\\
\label{eq:Eq-motion-xy-En}&\mathcal{E}^0=\sum\limits_k\frac32g(k)\sin^2\Phi(\varkappa)\left(\hat\psi_k\hat\psi_{-k}+\hat\psi_k^*\hat\psi_{-k}^*\right)\\ \nonumber
&+|\hat\psi_k|^2\left\{\ell^2k^2+[g(k)+2g(0)][2-3\sin^2\Phi(\varkappa)]\right\},
\end{align}
\end{subequations}
where function $\sin^2\Phi(\varkappa)$ is determined by Eq.~\eqref{eq:my-ground}.
The linear Eq.~\eqref{eq:Eq-motion-xy-Eq} has the solution $\hat\psi_k^\pm=\Psi_\pm e^{z_\pm(k)t}$, where the rate function $z_\pm(k)$ is
\begin{subequations}\label{eq:z-xy}
\begin{align}
\label{eq:rate-function}z_\pm(k)=\frac12\left[-\bar\varkappa\pm\sqrt{\bar\varkappa^2-4\alpha^+\alpha^-}\right],
\end{align}
where
\begin{align}\label{eq:alpha-pm}
\alpha^\pm(k)=&\ell^2k^2+[g(k)+2g(0)][2-3\sin^2\Phi(\varkappa)]\\ \nonumber
\pm&3g(k)\sin^2\Phi(\varkappa),\\
\bar\varkappa=&\frac{\varkappa}{2}\frac{1-\Lambda^{-2}}{1+\Lambda^{-2}}.
\end{align}
\end{subequations}
In accordance with Eq.~\eqref{eq:rate-function} the normalized current $\bar\varkappa$ plays a role of an effective damping\cite{*[{The effect of modifying of the spin-wave attenuation by the Slonczewski spin-torque in nanowires was also noted in }] [{}] XingAPL09}, that is why the natural damping can be omitted in the original equation \eqref{eq:LLS}. When
\begin{equation}\label{eq:instab-cond}
\alpha^+\alpha^-<0
\end{equation}
the rate $z_\pm$ becomes positive what results in instability of the stationary solution \eqref{eq:low-current-sol}. Analysis of \eqref{eq:alpha-pm} shows that for $h/\ell<\mathcal{C}_0$ with $\mathcal{C}_0\approx17.37$ the instability condition \eqref{eq:instab-cond} is equivalent to the condition $\sin^2\Phi>1/2$. Thus for the considered interval $-\pi/4\leq\Phi\leq0$ or, in other words, for the current interval $0\leq\varkappa\leq1/2$ the stationary solution \eqref{eq:low-current-sol} is stable. We do not analyze the case of very thick nanowires because the assumption of one-dimensionality of the magnetization does not work for the wire thickness $h/\ell\gg1$.

Thus, when the current is absent ($\varkappa=0$) the wire is magnetized uniformly along its axis. Adiabatically slow increasing of the spin-current from the value $\varkappa=0$ to value $\varkappa=1/2$ leads to the homogenous inclination of the wire magnetization by the angle $\Phi(\varkappa)$ which accordingly changes continuously from value $\Phi=0$ to value $\Phi=-\pi/4$, where the function $\Phi(\varkappa)$ is determined by \eqref{eq:low-current-sol}. The rotation takes place within the plane perpendicular to the current direction. The described regime takes place in region ``I'' and also in the hysteresis region ``I-III'' if the current is increasing, see Fig.~\ref{fig:mxy-vs-kappa} and Fig.~\ref{fig:diagram}. As one can see from the mentioned figures the analytically obtained behavior of the magnetization is in full agreement with micromagnetic simulations\footnote{We used the OOMMF code, version 1.2a5 [http://math.nist.gov/oommf/]. All simulations were performed for material parameters of permalloy: exchange constant $A = 1.3\times10^{-11}$~J/m, saturation magnetization $M_S = 8.6\times10^5$~A/m, and the anisotropy was neglected. The natural damping was also neglected $\alpha=0$ by the reasons explained in the text. Rate of polarization $\eta=0.4$ and $\Lambda=2$ were fixed for all simulations}.

The critical current $\varkappa_c=1/2$ in physical units reads
\begin{equation}\label{eq:Jc}
J_c=\frac{\pi M_s^2|e|h}{\hbar\eta}\left(1+\Lambda^{-2}\right).
\end{equation}
Thus dependence of the critical current $J_c$ on the wire thickness $h$ is linear one with the slope dependent on the parameter $\Lambda$. The corresponding dependencies $J_c(h)$ are shown in the Fig.~\ref{fig:diagram}a by lines 1, 1', and 1'' for different $\Lambda$-s.

\section{Stability of the saturated state. Periodic domain structure}\label{sec:domains}

\begin{figure}[h]
\includegraphics[width=\columnwidth]{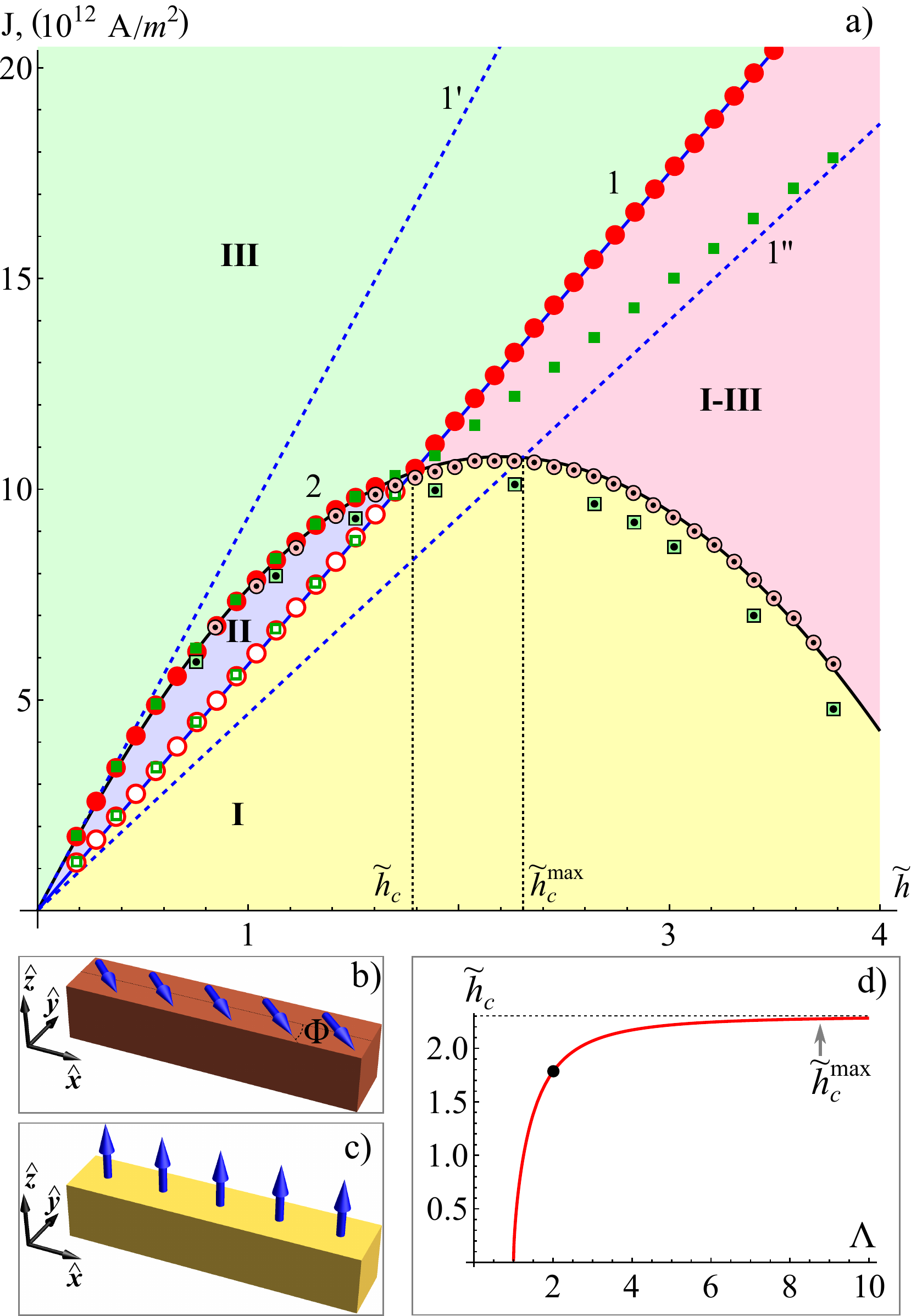}
\caption{Diagram of stationary states of permalloy square nanowires with the transverse size $h=\tilde h\ell$ under spin-polarized current influence, panel a). Sense of regions I, II, III and I-III is the same as in the Fig.~\ref{fig:mxy-vs-kappa}. Panels b) and c) show the magnetization states for regions I and II respectively. Lines 1, 1', and 1'' represent the critical current \eqref{eq:Jc} for the cases $\Lambda=2$, $\Lambda=1$, and $\Lambda\rightarrow\infty$ respectively. Line 2 shows the critical current \eqref{eq:Js}. Crossing of the lines 1 and 2 determines the critical thickness $\tilde h_c$ whose dependence on parameter $\Lambda$ is shown on the panel d), the point shows the parameters of simulations: $\Lambda=2$. Results of the micromagnetic simulations are shown by markers: disks and squares represent the cases of the periodic boundary conditions and finite wire length $L=1\,\mu m$ respectively; open and filled markers corresponds to transition from the inclined uniform state to the domain structure and to the saturated state respectively for the case of current increasing. Markers with dots shows the currents when the saturated state become instable for the case of the current decreasing.}\label{fig:diagram}
\end{figure}

\begin{figure}
\includegraphics[width=\columnwidth]{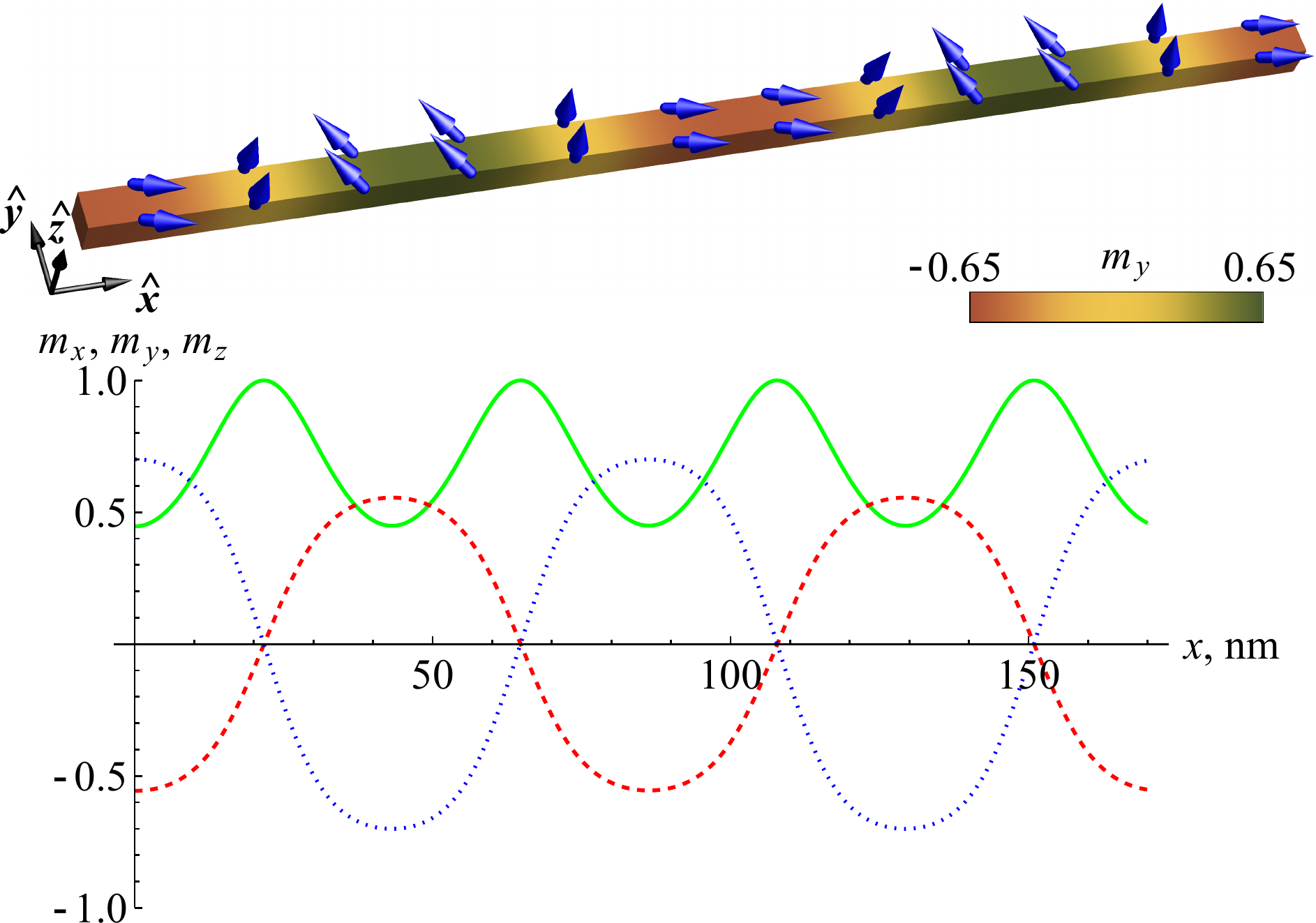}
\caption{The domain structure which appear in nanowires with transverse size $h<h_c$. The data obtained using micromagnetic simulations. The 170 nm length part of a nanowire with thickness $h=5$ nm and $L=1\mu m$ is shown in the upper part of the figure. Magnetization distribution of the domain structure is shown by arrows and color scheme. Magnetization components along the wire are also plotted at the bottom: solid, dashed and dotted lines correspond to $m_z$, $m_y$ and $m_x$ respectively.}\label{fig:domains}
\end{figure}

When the current overrides the value $J_c$ the inclined uniform state \eqref{eq:low-current-sol} becomes unstable. The new stationary state, which occurs as a result of the instability, depends on the wire thickness $h$. If $h$ exceeds some critical value $h_c$, which will be determined bellow, the nanowire very rapidly goes to the saturated regime where the magnetization is uniformly aligned along $\hat{\vec z}$-axis. This behavior is demonstrated in the Fig.~\ref{fig:mxy-vs-kappa}b.  It should be noted that contrary to infinite wires (or closed wires) the transition to saturation for wires of finite length is not quite a sharp jump due to the transitional formations of saturated domains at the wire ends. For thin wires with thickness $h<h_c$ the saturated state is unstable with respect to a periodic domain structure formation, see region ``II'' in the Figs.~\ref{fig:mxy-vs-kappa}, \ref{fig:diagram}. Magnetization distribution of the domain structure is shown in details in the Fig.~\ref{fig:domains}. Stability of the domain structure was proved using micromagnetic simulations.

To determine the critical thickness $h_c$ and the saturation current for $h<h_c$ we consider the stability of saturated state, when the quantization axis is directed along $\hat{\vec{z}}$-axis. Substituting $\Theta=0$ and $\Phi=0$\footnote{The angle $\Phi$ is not determined for the case $\Theta=0$, we chose $\Phi=0$ for a convenience.} into \eqref{eq:Ed-four-square} enable us to write the energy of the system in form
\begin{equation}
\begin{split}
\mathcal{E}^0=&\sum\limits_k|\hat\psi_k|^2\left[\ell^2k^2-g(k)-2g(0)\right]\\
-&\frac32g(k)\left(\hat\psi_k\hat\psi_{-k}+\hat\psi_k^*\hat\psi_{-k}^*\right)
\end{split}
\end{equation}
and equation of motion \eqref{eq:motion-four} in form
\begin{equation}\label{eq:Eq-motion-z}
-i\dot{\hat\psi}_k=\frac{\partial\mathcal{E}^0}{\partial\hat\psi^*_k}+i\tilde{\mathrm{j}}\hat\psi_k,
\end{equation}
where $\tilde{\mathrm{j}}=\mathrm{j}\eta/2$. Since the Eq.~\eqref{eq:Eq-motion-z} is linear we will look for solutions in form
\begin{equation}
\begin{split}
\hat\psi_{k}=\Psi_1^+ e^{\zeta_+(k)t}+\Psi_1^- e^{\zeta_-(k)t},\\
\hat\psi_{-k}^*=\Psi_2^+ e^{\zeta_+(k)t}+\Psi_2^- e^{\zeta_-(k)t}.
\end{split}
\end{equation}
In this case the rate function $\zeta_\pm(k)$ is
\begin{subequations}
\begin{align}
\label{eq:z-k}&\zeta_\pm(k)=-\tilde{\mathrm{j}}\pm\tilde\varkappa(k),\\
\label{eq:stab-function}&\tilde\varkappa(k)=\sqrt{9g^2(k)-\left[\ell^2k^2-g(k)-2g(0)\right]^2}.
\end{align}
\end{subequations}
As one can see from \eqref{eq:z-k} the current $\tilde{\mathrm{j}}$ plays the role of an effective damping as well as in case of the inclined solution \eqref{eq:low-current-sol}. For $\tilde{\mathrm{j}}>\tilde\varkappa(k)$ the rate function $\zeta_\pm<0$ and it means that the saturated state is stable. However, for  $\tilde{\mathrm{j}}<\max_k\tilde\varkappa(k)$ the rate function becomes negative $\zeta_+<0$ which means that the saturated state $\Theta=0$ is linearly unstable. Thus, the minimum current at which the saturated state is stable can be obtained as $\tilde{\mathrm{j}}_s=\max_k\tilde\varkappa(k)$ or equivalently in physical units
\begin{equation}\label{eq:Js}
J_s=\frac{8\pi M_s^2|e|h}{\hbar\eta}\max\limits_k\tilde\varkappa(k)
\end{equation}
Dependence of the critical current \eqref{eq:Js} on the transverse wire size $h$ is shown by line 2 in the diagram of states, see Fig.~\ref{fig:diagram}a. The dependence $J_s(h)$ is not linear due to the nontriviality of dependence $\tilde\varkappa(h)$, see \eqref{eq:stab-function} and \eqref{eq:g-approx}.

It is important to note that the appearance of the periodic domain structure essentially depends on the resistance mismatch parameter $\Lambda$. Contrary to $J_c$ the critical current $J_s$ does not depend on the parameter $\Lambda$. Using \eqref{eq:g-approx} and \eqref{eq:stab-function}, one can easily show that $\max\limits_k\tilde\varkappa(k)\rightarrow3g(0)=1/4$ for the case $h\rightarrow0$. Comparing now \eqref{eq:Js} and \eqref{eq:Jc} we conclude that for $\Lambda=1$ (minimal possible value) the line $J_c(h)$ is tangent to the curve $J_s(h)$ in point $h=0$, see line 1' in the Fig.~\ref{fig:diagram}a. In this case $J_c(h)>J_s(h)$ and this means that the saturation, which appears when the current overrides the value $J_c$, remains stable and the domain structure is not formed. For the other case $\Lambda>1$ the slope of the line $J_c(h)$ decreases and the intersection of the dependencies $J_c(h)$ and $J_s(h)$ determines the critical thickness $h_c$:
\begin{equation}\label{eq:h_c}
J_c(h_c)=J_s(h_c)
\end{equation}
such that for $h<h_c$ the saturated state is unstable for currents $J<J_s$ and the stable periodic domain structure appears in the interval $J_c<J<J_s$.

For $h>h_c$ we have $J_c(h)>J_s(h)$ and a hysteresis takes place. When current increases from zero and reaches the critical value $J_c$ the inclined uniform state \eqref{eq:low-current-sol} becomes unstable and the system abruptly goes to the saturated state. By moving in opposite direction, i.e. when we start to decrease current from the saturated state it remains stable down to the current $J_s<J_c$. Note in passing that the transition from the saturated state to the inclined state \eqref{eq:low-current-sol} may be accompanied by appearance of an irregular domain structure which arises due to degeneracy of the inclined state \eqref{eq:low-current-sol}: if $\Phi=\Phi_0$ is a solution of \eqref{eq:low-current-sol} then $\Phi=\Phi_0+\pi$ is another solution. The described hysteresis region is denoted ``I-III'' in the Figs.~\ref{fig:mxy-vs-kappa}b), and \ref{fig:diagram}a).

Thus, under the adiabatic increase of the current the saturation value of the current depends on the wire thickness:
\begin{equation}\label{eq:J-satur}
\mathfrak{J}=\begin{cases}&J_s,\quad h<h_c,\\
&J_c,\quad h>h_c.
\end{cases}
\end{equation}
It is worth noting that for $h>h_c$ the saturation current for wires of finite length is slightly lower than $J_c$, see Fig.~\ref{fig:diagram}a. This happens due to a particular role of the wire ends in the process of the saturation: for the currents $J\lessapprox J_c$ domains magnetized along $\hat{\vec z}$-axis occur at each of the ends. The length of this end domains rapidly increases with the current increasing and eventually the wire goes to the saturation state when the length of each of the domains reaches the half of the wire length.

The critical thickness $h_c$ determined by \eqref{eq:h_c} depends on the parameter $\Lambda$, see Fig.~\ref{fig:diagram}d. With $\Lambda$ increasing the quantity $h_c$ asymptotically approaches its maximum value $h_c^{\mathrm{max}}\approx2.303\ell$, see line 1'' in the Fig.~\ref{fig:diagram}a. For the minimal possible value $\Lambda=1$ the critical thickness $h_c=0$.

\section{Conclusions}
We study theoretically the influence of perpendicular spin-polarized current on magnetization behavior of the long ferromagnetic nanowire of square cross-section. We found out that under the action of current the wire always reaches some stationary state. This is in contrast to the case of planar films where either dynamical regimes or stationary states may realize for different current densities\cite{Volkov11,Gaididei12a}. In the no current case the wire is uniformly magnetized along its axis. For current increasing within the interval $0<J<J_c$ the wire magnetization remains spatially uniform being inclined in the plane perpendicular to the current direction. The inclination angle $\Phi$ depends on the current density $J$. The critical current $J_c$ depends on the thickness and the mismatch parameter $\Lambda$. When the current overrides the value $J_c$ the nanowire goes either to the saturated state where the magnetization is uniformly aligned along the direction of spin-polarization of the current or to a periodic multidomain state, depending on the wire thickness $h$. The domain structure exists for $h<h_c$ in the interval of currents $J_c<J<J_s$. In contrast to $J_c$ the upper value of this interval $J_s$ does not depend on the resistance mismatch parameter $\Lambda$.

For thick wires with $h>h_c$ an opposite inequality $J_c>J_s$ holds and a hysteresis phenomenon takes place.


\appendix

\section{Equation of motion in terms of $\psi$-function}\label{ap:the-equation}
In accordance to \eqref{eq:tyablikov} one can consider that $\vec m_n=\vec m_n(\psi,\,\psi^*)$. Projecting the Eq.~\eqref{eq:LLS} to axes $x$ and $y$ respectively results in the system
\begin{equation}\label{eq:system-mx-my}
\begin{split}
&\frac{\partial m^x}{\partial\psi}\dot\psi+\frac{\partial m^x}{\partial\psi^*}\dot\psi^*=m^y\frac{\partial\mathcal{E}}{\partial m^z}-m^z\frac{\partial\mathcal{E}}{\partial m^y}-j\varepsilon m^x m^z,\\
&\frac{\partial m^y}{\partial\psi}\dot\psi+\frac{\partial m^y}{\partial\psi^*}\dot\psi^*=m^z\frac{\partial\mathcal{E}}{\partial m^x}-m^x\frac{\partial\mathcal{E}}{\partial m^z}-j\varepsilon m^y m^z,
\end{split}
\end{equation}
where we omitted index $n$ for the sake of simplicity. Solving the system \eqref{eq:system-mx-my} with respect to $\dot\psi$ and $\dot\psi^*$ one obtains
\begin{subequations}\label{eq:eq-motion-psi-pre}
\begin{align}
\label{eq:psi-dot-pre}&\dot\psi=-\frac{m^z}{D}\left[\frac{\partial\mathcal{E}}{\partial\psi^*}+i\mathcal{F}^{st}\right],\\
\label{eq:Fst-pre}&\mathcal{F}^{st}=-ij\varepsilon\left(m^x\frac{\partial m^y}{\partial\psi^*}-m^y\frac{\partial m^x}{\partial\psi^*}\right).
\end{align}
\end{subequations}
Here $D$ is the determinant of the system \eqref{eq:system-mx-my}:
\begin{equation}\label{eq:determinant}
D=\frac{\partial m^x}{\partial \psi}\frac{\partial m^y}{\partial \psi^*}-\frac{\partial m^x}{\partial \psi^*}\frac{\partial m^y}{\partial \psi}.
\end{equation}
Substitution \eqref{eq:m-components} into \eqref{eq:determinant} results in $D=im^z$ and finally the equation \eqref{eq:eq-motion-psi-pre} takes the form of \eqref{eq:eq-motion-psi} with spin-torque term \eqref{eq:Fst-psi-exact}.
\section{Dipole-dipole contribution}\label{ap:dipole}
Substituting \eqref{eq:m-components} into \eqref{eq:Ed-1D}, we obtain that the harmonic part of the dipole-dipole energy in the wave-vector space in form of
\begin{equation}\label{eq:Ed-four}
\begin{split}
\mathcal{E}^0_d=&-\sum\limits_k\biggl\{3\sqrt{2\mathcal{N}_x}\sin\Theta\hat\alpha_0(0)\hat\psi_k+\frac32\hat\alpha_1(k)\hat\psi_k\hat\psi_{-k}\\
+&\frac12|\hat\psi_k|^2\left[\hat\alpha_2(k)+2\hat\alpha_2(0)\right]+c.c.\biggr\},
\end{split}
\end{equation}
where the coefficients $\hat\alpha_i(k)$ are
\begin{widetext}
\begin{equation}\label{eq:alphas}
\begin{split}
&\hat\alpha_0(k)=\frac{a^3}{8\pi\mathcal{N}_s}\sum\limits_n\sum\limits_{\bar{\vec\nu},\bar{\vec\mu}}\frac{\cos\Theta\left(n^2\cos^2\Phi+y_{\vec\nu\vec\mu}^2\sin^2\Phi-z_{\vec\nu\vec\mu}^2\right)+i\sin\Phi\cos\Phi\left(n^2-y_{\vec\nu\vec\mu}^2\right)} {\left(n^2+y_{\vec\nu\vec\mu}^2+z_{\vec\nu\vec\mu}^2\right)^{5/2}}e^{ikn},\\
&\hat\alpha_1(k)=\frac{a^3}{8\pi\mathcal{N}_s}\sum\limits_n\sum\limits_{\bar{\vec\nu},\bar{\vec\mu}}\frac{\left(n^2-y_{\vec\nu\vec\mu}^2\right)\left(\cos2\Phi+i\cos\Theta\sin2\Phi\right)-\sin^2\Theta\left(n^2\cos^2\Phi+y_{\vec\nu\vec\mu}^2\sin^2\Phi-z_{\vec\nu\vec\mu}^2\right)} {\left(n^2+y_{\vec\nu\vec\mu}^2+z_{\vec\nu\vec\mu}^2\right)^{5/2}}e^{ikn},\\
&\hat\alpha_2(k)=\frac{a^3}{8\pi\mathcal{N}_s}\sum\limits_n\sum\limits_{\bar{\vec\nu},\bar{\vec\mu}}\frac{n^2+y_{\vec\nu\vec\mu}^2-2z_{\vec\nu\vec\mu}^2-3\sin^2\Theta\left(n^2\cos^2\Phi+y_{\vec\nu\vec\mu}^2\sin^2\Phi-z_{\vec\nu\vec\mu}^2\right)} {\left(n^2+y_{\vec\nu\vec\mu}^2+z_{\vec\nu\vec\mu}^2\right)^{5/2}}e^{ikn}.
\end{split}
\end{equation}
\end{widetext}
For the wires with cross-sections in form of square, disk or ring the indexes $\nu_y$ and $\nu_z$ are interchangeable, as well as indexes $\mu_y$ and $\mu_z$. That enables us to to simplify the expressions \eqref{eq:alphas}
\begin{equation}
\begin{split}
&\hat\alpha_0(k)=\cos\Phi(\cos\Theta\cos\Phi+i\sin\Phi)g(k),\\
&\hat\alpha_1(k)=(\cos2\Phi-\sin^2\Theta\cos^2\Phi+i\cos\Theta\sin2\Phi)g(k),\\
&\hat\alpha_2(k)=(1-3\sin^2\Theta\cos^2\Phi)g(k),
\end{split}
\end{equation}
where function $g(k)$ is the following
\begin{equation}\label{eq:g-sum}
g(k)=\frac{a^3}{8\pi\mathcal{N}_s}\sum\limits_n\sum\limits_{\bar{\vec\nu},\bar{\vec\mu}}\frac{n^2-y_{\vec\nu\vec\mu}^2}{\left(n^2+y_{\vec\nu\vec\mu}^2+z_{\vec\nu\vec\mu}^2\right)^{5/2}}e^{ikn}.
\end{equation}

\begin{figure}
\includegraphics[width=0.8\columnwidth]{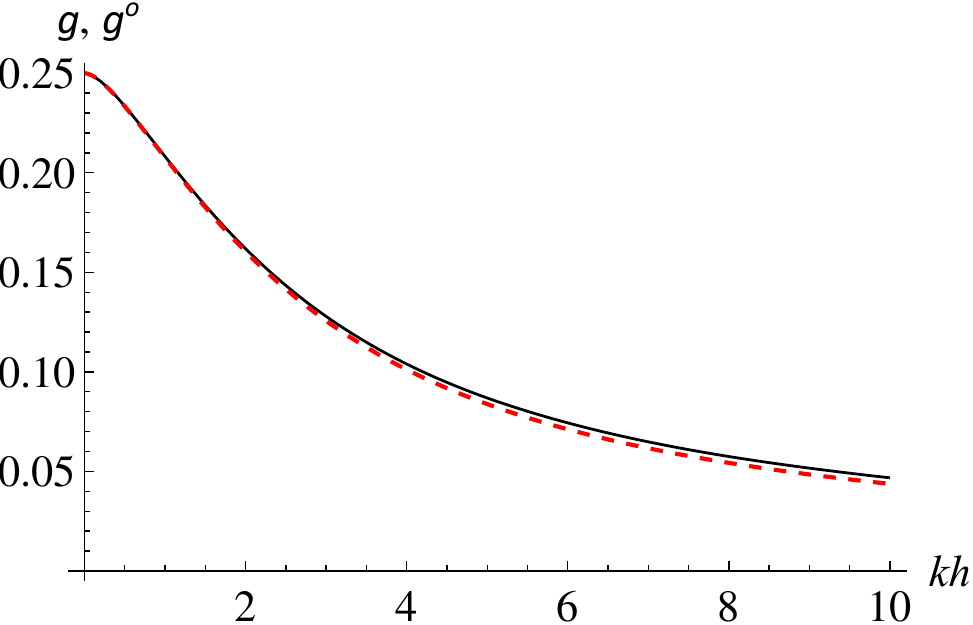}
\caption{Comparison of regular parts of sum \eqref{eq:g-sum} for wires with square and round cross-sections. The solid and dashed lines correspond to the square and round wires respectively and represent expressions \eqref{eq:g-reg-square} and \eqref{eq:g-round-fin} respectively.}\label{fig:g-functions}
\end{figure}

To obtain more suitable expression for function $g(k)$ we proceed in \eqref{eq:g-sum} from summation to integration. In this case $g(k)$ can be presented as sum of regular and singular parts in the following way
\begin{subequations}
\begin{align}
\label{eq:g-main}g(k)&=g_{reg}(k)+g_{sing}(k),\\
\label{eq:g-reg}g_{reg}(k)&=\frac{1}{\pi h^2}\lim\limits_{r_0\rightarrow0}\int\limits_{r_0}^{\infty}\mathrm{d}x\cos(kx)\int\limits_{0}^h\mathrm{d}y\int\limits_{0}^h\mathrm{d}z\times\\\nonumber
&\times(h-y)(h-z)\frac{x^2-y^2}{\left[x^2+y^2+z^2\right]^{5/2}},\\
\label{eq:g-sing}g_{sing}(k)&=\frac{1}{\pi h^2}\lim\limits_{r_0\rightarrow0}\int\limits_{0}^{r_0}\!\mathrm{d}x\cos(kx)\!\!\int\limits_0^{\pi/2}\!\!\mathrm{d}\chi\!\!\!\!\!\int\limits_{\sqrt{r_0^2-x^2}}^{\sigma(\chi)}\!\!\!\!\!\mathrm{d}rr\times\\\nonumber
&\times(h-r\sin\chi)(h-r\cos\chi)\frac{x^2-r^2\cos^2\chi}{\left(x^2+r^2\right)^{5/2}}.
\end{align}
\end{subequations}
Here we used the relation
\begin{equation}
\int\limits_0^h\mathrm{d}y\int\limits_0^h\mathrm{d}y'F(|y-y'|)=2\int\limits_0^h\mathrm{d}y(h-y)F(y).
\end{equation}
Furthermore in \eqref{eq:g-sing} we proceed to the polar frame of reference $(r,\chi)$ originated in one of the vertexes of the wire cross-section, and function $\sigma(\chi)$ determines the cross-section shape. In the following we will ascertain that specific form of the function $\sigma(\chi)$ does not affect the value of the integral \eqref{eq:g-sing}. To do this one should change the variables $r=r_0\rho$, $x=r_0\xi$ and find the limit in \eqref{eq:g-sing}. The result is the following
\begin{equation}\label{eq:g-sing-pre}
g_{sing}(k)=\frac{1}{\pi}\int\limits_0^1\mathrm{d}\xi\int\limits_0^{\pi/2}\mathrm{d}\chi\!\!\!\int\limits_{\sqrt{1-\xi^2}}^\infty\!\!\!\mathrm{d}\rho\rho\frac{\xi^2-\rho^2\cos^2\chi}{(\xi^2+\rho^2)^{5/2}}.
\end{equation}
The direct integration of \eqref{eq:g-sing-pre} results in
\begin{equation}\label{eq:g-sing-fin}
g_{sing}(k)=-\frac16.
\end{equation}

Let us now proceed to the calculation of the regular part \eqref{eq:g-reg}. The direct integration over variables $y$ and $z$ with the next integration by parts over $x$ enables us to simplify expression \eqref{eq:g-reg}
\begin{equation}\label{eq:g-reg-square}
\begin{split}
g_{reg}(k)=&\frac{1}{\pi kh}\int\limits_0^\infty\sin(khx)\biggl(\frac{\sqrt{1+x^2}}{x}+\frac{x}{\sqrt{1+x^2}}\\
-&x\frac{\sqrt{2+x^2}}{1+x^2}-1\biggr)\mathrm{d}x.
\end{split}
\end{equation}
Using the representation \eqref{eq:g-reg-square} one can easily obtain $g_{reg}(0)=1/4$ and consequently
\begin{equation}
g(0)=\frac{1}{12}.
\end{equation}
Let us consider the case $hk\ll1$. In other words, the wire thickness is much smaller then the wavelength, it means that the form of the wire cross-section should not be of principle. To show this we calculate the regular part of function $g(k)$ for a nanowire with disk-shaped cross-section with radius $R=h/\sqrt\pi$, so the cross-section areas of round and square wires are equal. In this case the regular part of the sum \eqref{eq:g-sum} can be presented by integral
\begin{equation}\label{eq:g-round}
\begin{split}
g_{reg}^o(k)=&\frac{1}{4\pi h^2}\lim\limits_{r_0\rightarrow0}\int\limits_{r_0}^\infty\!\mathrm{d}x\!\!\int\limits_0^{2\pi}\!\!\mathrm{d}\chi\!\!\int\limits_0^{2\pi}\!\!\mathrm{d}\chi'\!\!\!\!\int\limits_0^{h/\sqrt\pi}\!\!\!\!\mathrm{d}r\!\!\!\!\int\limits_0^{h/\sqrt\pi}\!\!\!\!\mathrm{d}r'\\
\times&\frac{rr'\cos(kx)\left[x^2-(r\cos\chi-r'\cos\chi')^2\right]}{\left[x^2+r^2+r'^2-2rr'\cos(\chi-\chi')\right]^{5/2}}
\end{split}
\end{equation}
Using the parametrization
$$
3\frac{p^2}{(p^2+c^2)^{5/2}}-\frac{1}{(p^2+c^2)^{3/2}}=\int\limits_0^\infty\xi^2e^{-p\xi}J_0(c\xi)\mathrm{d}\xi
$$
and relation
$$
\int\limits_0^{2\pi}J_0(\sqrt{r^2+r'^2-2rr'\cos\chi})=2\pi J_0(r)J_0(r'),
$$
where $J_0(x)$ is Bessel Function of the first kind, one can easily integrate the expression \eqref{eq:g-round} and obtain
\begin{equation}\label{eq:g-round-fin}
g_{reg}^o(k)=\frac12I_1\left(\frac{kh}{\sqrt\pi}\right)K_1\left(\frac{kh}{\sqrt\pi}\right),
\end{equation}
where $I_1(x)$ and $K_1(x)$ are modified Bessel functions of the first and second kinds respectively.
Since the functions $g_{reg}(k)$ and $g_{reg}^o(k)$ are very close, see Fig.~\ref{fig:g-functions}, one can use the approximation \eqref{eq:g-approx}.


%

\end{document}